\documentclass[review]{elsarticle}
\usepackage{lineno,hyperref}
\modulolinenumbers[5]

\journal{Elsevier}

\usepackage{tabularx}
\usepackage{graphicx}
\usepackage{latexsym}
\usepackage{color}
\usepackage{caption}
\usepackage{subcaption}
\usepackage{bm}
\usepackage{bbm}
\usepackage{amsmath,amssymb, amsthm}
\usepackage{algorithm}
\usepackage{algpseudocode}
\usepackage{color,soul}
\usepackage{graphics}
\usepackage{amsfonts}
\usepackage{setspace}   
\usepackage{wasysym}
\usepackage{multirow}

\usepackage{adjustbox}
\usepackage{graphicx}
\usepackage{multirow}
\usepackage{nomencl}
\usepackage{float}
\usepackage[table]{xcolor}% http://ctan.org/pkg/xcolor}
\makenomenclature
\newcommand*\diff{\mathop{}\!\mathrm{d}}
\usepackage{geometry}
\newtheorem{theorem}{Theorem}

\theoremstyle{definition}

\begin{document}

\title{A preconditioning approach for improved estimation of \\ sparse polynomial chaos expansions}
\author{Negin Alemazkoor}
\author{Hadi Meidani}
\address{Department of Civil and Environmental Engineering, University of Illinois at Urbana-Champaign, Urbana, Illinois, USA.}

\begin{abstract}
Compressive sampling has been widely used for sparse  polynomial chaos (PC) approximation of stochastic functions. The recovery accuracy of compressive sampling highly depends on the incoherence properties of the measurement matrix.  In this paper, we consider preconditioning the underdetermined system of equations that is to be solved. Premultiplying a linear equation system by a non-singular matrix results in an equivalent equation system, but it can potentially improve the incoherence properties of the resulting preconditioned measurement matrix and lead to a better recovery accuracy. When measurements are noisy, however, preconditioning  can also potentially result in a worse signal-to-noise ratio, thereby deteriorating recovery accuracy.  In this work, we propose a preconditioning scheme that improves the incoherence properties of measurement matrix and at the same time  prevents undesirable deterioration of signal-to-noise ratio. We provide theoretical motivations and numerical examples that demonstrate the promise of the proposed approach in improving the  accuracy of  estimated polynomial chaos expansions.
\end{abstract} 
\maketitle
%%\tableofcontents

\section{Introduction} \label{sec:intro}
Reliable analysis of natural and engineered systems necessitates the understanding of how the system response or  quantity of interest (QoI) depends on uncertain inputs. Uncertainty quantification (UQ) tackles these issues with efficient propagation of input uncertainties onto the QoI. This is typically done by building approximate surrogates that replace computationally expensive simulations. Surrogates approximate QoI as an analytical function of random inputs and facilitate quantifying parametric uncertainty. As a widely used spectral surrogate, the  polynomial chaos expansion (PCE) uses orthogonal polynomials for the approximation of QoI. In order to construct PCEs, i.e., to determine the expansion coefficients corresponding to different polynomial bases, a widely used approach is stochastic collocation. This approach is advantageous particularly because it is nonintrusive and easy to implement  by reusing legacy codes. Examples of stochastic collocation methods include spectral projection \cite{conrad2013adaptive,constantine2012sparse},  sparse grid interpolation \cite{barthelmann2000high,buzzard2012global}, and least square \cite{shin2016nonadaptive,hampton2015coherence}. The outstanding challenge is that the number of  sample points, e.g. sampled simulations, that these methods require for accurate prediction increases drastically as the number of uncertain inputs, or the dimension of the PCE, increases.

Recently, motivated by the fact that approximated PCEs for many high dimensional problems could be sparse, compressive sampling has been effectively used to approximate PCE coefficients using  a small number of sampled simulations   and has thus alleviated the dimensionality-related challenge\cite{doostan2011non,guo2017stochastic,yan2012stochastic,alemazkoor2017near,peng2014weighted,yan2017sparse}. The accuracy of sparse polynomial approximation heavily relies on incoherence properties of measurement matrix, which is a matrix that consists of evaluated polynomial bases at sample points. Most commonly,  in standard sampling, samples are generated randomly from the probability distribution of random inputs. Recently, other sampling strategies have been proposed in order to improve the accuracy of sparse PCEs. In \cite{rauhut2012sparse}, it was proposed to draw samples from Chebyshev probability distribution to construct a sparse Legendre-based PCE. The theoretical and numerical results in \cite{yan2012stochastic} showed that  Chebyshev sampling deteriorates the recovery accuracy  in high dimensional problems. Drawing samples randomly from the tensor grid of Gaussian quadrature points was recommended in \cite{guo2017stochastic, tang2014subsampled}. Although the results showed significant accuracy improvement in low-dimensional problems, the results underperformed or were close to standard sampling in  high-dimensional problems. A sampling strategy that outperforms standard sampling in both low-dimensional high-order and high-dimensional low-order problems was proposed in \cite{hampton2015compressive}, and was called the coherence-optimal sampling. In coherence-optimal sampling, samples are drawn from a measure that minimizes the (local) coherence of orthogonal polynomial system. In \cite{alemazkoor2017near}, a near-optimal sampling strategy was proposed, which further improved coherence-optimal sampling by subsampling a few samples from a large pool of coherence-optimal samples in such a way that the resulting measurement matrix will have better cross-correlation properties. 

In this work, instead of relying on experimenting with sampling strategies to improve  incoherence properties, we take an alternative approach and use a preconditioning scheme to enhance these properties. The term \textit{preconditioning} has already been used in the sparse PCE  literature to refer to a weight matrix that is added to preserve asymptotic orthogonality of  basis functions when non-standard sampling distributions are used \cite{jakeman2017generalized, hampton2015compressive}. In underdetermined cases where a limited set of samples are available, enforcing asymptotic orthogonality does not necessarily guarantee  improvement in  incoherence properties of actual measurement matrices. 

To the best of our knowledge, our present work is the first attempt at preconditioning the underdetermined measurement matrix towards better incoherence properties for  polynomial chaos approximation. When the target stochastic function is exactly sparse with respect to the truncated PC bases and measurements are noiseless, one can borrow methods for designing optimal projection (or sensing) matrix in signal processing applications \cite{abolghasemi2012gradient,li2013projection} to precondition underdetermined equation systems. This is because both the preconditioning matrix and the projection matrix will appear in the same way in the coherence-based objective that one seeks to optimize. However, when modeling or truncation error or measurement noise is present, that  resemblance no longer exists. Therefore, in these polynomial regression problems, an improper design of preconditioning matrix can undesirably amplify the noise vector more than it does the measurement vector. This can result in a worse "signal-to-noise" ratio leading to inaccurate polynomial approximation.

 In this work, we  propose an original approach for designing the preconditioning matrix such that (1) the incoherence properties of the resulting preconditioned measurement matrix are improved, and (2) undesirable amplification of noise vector vis-\'a-vis measurement vector is prevented. The proposed preconditioning technique is also capable of preconditioning a system with no measurement or truncation error, and as such has general applicability. Using numerical examples, it will be demonstrated that preconditioning can significantly improve the accuracy of  PCE surrogates built using compressive sampling. This paper is organized as follows: Section \ref{sec:CS in UQ} provides a brief overview on application of compressive sampling in PCE approximation; Section \ref{sec:preconditioning}  introduces the preconditioning scheme along with the theoretical motivation, and Section  \ref{sec:Numerical examples} includes numerical illustration and detailed discussion about the comparative performance of the proposed preconditioning scheme. 
  
\section{Setup and background} \label{sec:CS in UQ}
\subsection{Polynomial chaos expansion}
Consider the vector of independent random variables $\bm \Xi=(\Xi_1, ..., \Xi_d)$ to be $d$-dimensional system random inputs and $u(\bm \Xi)$ to be the uncertain QoI with finite variance. Then, $u(\bm \Xi)$ can be written as an expansion of orthogonal polynomials, i.e.
\begin{equation}\label{eq:repUasSum}
u(\bm{\Xi})= \sum_{\bm \alpha \in \mathbb{N}_0^d} c_{\bm \alpha } \psi_{\bm \alpha}(\bm\Xi). 
\end{equation}
Let $I_{\bm \Xi} \subseteq \mathbb{R}^d $ be a tensor-product domain that is the support of $\bm \Xi$, i.e. $\Xi_i \in I_{ \Xi_i}$ and $I_{\bm \Xi} = \times_{i=1}^d I_{ \Xi_i} $. Also, let $\rho_i: I_{\Xi_i} \rightarrow \mathbb{R}^+$ be the probability measure for variable $\Xi_i$ and let $\rho(\bm \Xi)=\prod_{i=1}^{d}\rho_i(\Xi_i)$. Given this setting, the set of univariate orthonormal polynomials, $\{ \psi_{\alpha,i} \}_{ \alpha \in \mathbb{N}_0}$ , satisfies
\begin{equation} \label{eq:orthonormality1d}
\int_{I_{\Xi_i}} \psi_{ n,i}(\xi_i) \psi_{ m,i}(\xi_i)  \rho_i({\xi_i}) \diff  \xi_i = \delta_{{mn}}, \quad {m,n} \in \mathbb{N}_0.
\end{equation}
Consequently, the $d$-dimensional orthonormal polynomials $\{ \psi_{\bm \alpha} \}_{\bm \alpha \in \mathbb{N}_0^d}$ in $\bm \Xi$ are formed by the product of the univariate orthonormal polynomials, 
\begin{equation}
\psi_{\bm \alpha} (\bm \Xi) = \prod_{i=1}^{d} \psi_{\alpha_i,i} (\Xi_i),     
\end{equation}
where $\alpha_i$ represents the $i$th coordinate of $\bm \alpha$. For practicality, we need to truncate the expansion in (\ref{eq:repUasSum}) by limiting the total order of polynomial to be $k$ and only including bases with $\left \| \bm \alpha \right \|_1 \leqslant k$ in the expansion. The cardinality of basis set, denoted by $K$, will then be
\begin{equation}\label{eq:NumOfbasis}
K:= \frac{(k+d)!}{k! \ d!}.
\end{equation}
The resulting  truncated PCE,  $u_k(\bm \Xi)$, offers a practical approximation of $u(\bm \Xi)$ (inducing a truncation error),
\begin{equation}\label{eq:Approximation}
u(\bm\Xi)   \approx u_{k}(\bm\Xi) := \sum_{\left \| \bm \alpha \right \|_1 \leqslant k} c_{\bm \alpha } \psi_{\bm \alpha}(\bm \Xi).
 \end{equation}
The exact PCE coefficients in Equation~\ref{eq:repUasSum} can be exactly calculated by projecting $u(\bm \Xi)$ onto the basis functions $\psi_{\bm \alpha}$:
\begin{equation}\label{eq:exactCoeff}
c_{\bm \alpha} = \mathbb{E}_{\rho}\begin{bmatrix}
u(\bm \Xi) \psi_{\bm \alpha}(\bm \Xi)
\end{bmatrix}= \int_{I_{\bm \Xi}}u(\bm \xi)\psi_{\bm \alpha}(\bm \xi) \rho(\bm \xi) \diff \bm \xi. 
\end{equation}
In the typical absence of analytical solution for this multidimensional integral, the PCE coefficients  can be numerically calculated using non-intrusive approaches such as Monte Carlo sampling and sparse grid quadrature \cite{xiu2010numerical}. However, it is known that Monte Carlo sampling converges slowly and sparse grid quadrature can be impractical in high-dimensional problems \cite{xiu2010numerical,bungartz2004sparse}. In practice, for high-dimensional problems, least squares regression is widely used for the calculation of PCE coefficients. This approach requires the number of samples to be larger than the number of unknown coefficients, in order to make the system of equations overdetermined. The generally accepted oversampling rate is about 1.5 to 3 times the number of coefficients  \cite{shin2016near, shin2016nonadaptive}. Recently, a quasi-optimal sampling approach was introduced in \cite{shin2016nonadaptive} that allows accurate coefficient recovery with $\mathcal{O} (1)$ oversampling rate. However, affording that many samples may still be computationally impossible. In \cite{doostan2011non} it was shown that when the QoI is sparse with respect to the PCE bases, compressive sampling can be used to calculate the expansion coefficients using samples that are much fewer than the unknown coefficients. This section follows with a brief introduction of compressive sampling application in PCE approximation. 
\subsection{Sparse PCE estimation via compressive sampling} \label{recovery}
Compressive sampling  first emerged in the field of signal possessing and has since  found applications in various domains, such as radar systems \cite{ender2010compressive}, speech recognition \cite{gemmeke2010compressive} and MR imaging \cite{lustig2007sparse}. Compressive sampling solves an underdetermined system of equations by exploring its sparsest solution. Here, we present the  compressive sampling in the context of stochastic collocation. Specifically, consider $M$ realizations  $\left \{ \bm \xi^{(i)} \right \}_{i=1}^M$ of system inputs, with corresponding model outputs $\bm u=(u (\bm \xi^{(1)}),...,u (\bm \xi^{(M)}))^T$. We seek a solution that satisfies 
\begin{equation}\label{Axb}
\bm \Psi \bm c= \bm u,
\end{equation}
where  $\bm \Psi$ is the Vandermonde-like matrix, often referred to as the measurement or design matrix, and is constructed according to 
\begin{equation} \label{mesmatrix}
\bm \Psi=[\psi_{ij}], \quad \psi_{ij}=\psi_{\bm \alpha^{j}}(\bm \xi ^{(i)}), \quad 1\leqslant i\leqslant M, \quad  1\leqslant j\leqslant K. 
\end{equation}
The problem ~\ref{Axb} is underdetermined and ill-posed when $M<K$. Therefore, obtaining a unique solution requires adding  regularization to the problem. In compressive sampling, the main goal is to determine the sparsest solution, which can be achieved by minimizing the $\ell_0$-norm of the solution according to 
\begin{equation} \label{eq:l0min}
\underset{\bm c}{\textrm{min}} \left \| \bm c \right \|_{0} \quad \textrm{subject to} \quad  \bm \Psi \bm c= \bm u.
\end{equation}
However, since the $\ell_0$-norm is non-convex and discontinuous, the above problem is NP-hard. Therefore, $\ell_0$~minimization is usually replaced with its convex relaxation, where the $\ell_1$-norm of the solution $\bm c$ is minimized instead, i.e.,
\begin{equation} \label{eq:l1min}
\underset{\bm c}{\textrm{min}} \left \| \bm c \right \|_{1} \quad \textrm{subject to} \quad  \bm \Psi \bm c= \bm u.
\end{equation}
Under certain conditions, $\ell_1$ minimization gives the same solution as $\ell_0$ minimization \cite{candes2006stable}. The $\ell_1$ minimization problem is commonly known as the \textit{Basis Pursuit}. In case of noisy measurements or when $u(\bm \Xi)$ can not be exactly represented by $k$th order PC basis, one can use the \textit{Basis Pursuit Denoising} \cite{chen2001atomic} formulated as
\begin{equation} \label{eq:l1minNo}
\underset{\bm c}{\textrm{min}} \left \| \bm c \right \|_{1} \quad \textrm{subject to} \quad  \left \|\bm \Psi \bm c- \bm u \right\|_{2} \leq \epsilon,
\end{equation}
where  $\epsilon$ is the error tolerance.  Other types of regularizations such as \textit{Least Absolute Shrinkage and Selection Operator (lasso)}, $\ell_{1-2}$ minimization and $\ell_{0.5}$ regularization can also be used \cite{ tibshirani1996regression,tibshirani2011regression, yin2015minimization,xu20101}.
\subsection{$\ell_1$ minimization recoverability}
The ability of $\ell_1$ minimization in accurate solution recovery depends on the actual sparsity of the solution $\bm c$ and properties of the measurement matrix $\bm \Psi$. Here, we discuss  one of the main properties of $\bm \Psi$, shown to be significantly relevant to recovery accuracy. 

\newtheorem{RIPdef}{Definition}
\begin{RIPdef} (restricted isometry constant \cite{candes2008restricted})\textit{The $s$-restricted isometry constant for a matrix $\bm \Psi \in \mathbb{R}^{M \times K}$ is defined to be the smallest $\delta_{s} \in (0,1)$ such that
	\begin{equation} \label{eq:RIP}
	(1-\delta_{s})\left \| \bm c \right \|_{2}^{2}\leqslant  \left \| \bm \Psi \bm c\right \|_{2}^{2}\leqslant  (1+\delta_{s})\left \| \bm  c \right \|_{2}^{2},
	\end{equation}
	for all $\bm c \in \mathbb{R}^{K}$ that are at most $s$-sparse.} 
\end{RIPdef}
Depending on the property  of this isometry constant, one can expect  the $\ell_1$  minimization (\ref{eq:l1min}) to produce a very accurate and even exact solution, as stated in the following theorem.  
\begin{theorem}[\cite{candes2008restricted}]\label{thm.rip}
	Let $\bm \Psi \in \mathbb{R}^{M\times K}$ with isometry constant $\delta_{2s}$ such that $\delta_{2s}<\sqrt{2}-1$. For a given $ \bar{\bm c}$, and measurement $\bm y=\bm\Psi \bar{\bm c}$,  let $\bm c$ be the solution of 
	\begin{equation} \label{eq:l1minerror1}
	{\textrm{min}}\left \| \bm c \right \|_{1} \quad \textrm{subject to} \quad  \bm \Psi \bm c = \bm y.
	\end{equation}
	Then the reconstruction error satisfies 
	\begin{equation}
	\left \| \bm c-\bar{\bm c} \right \|\leqslant C_1\frac{\left \| \bar{\bm c}- \bm c^{*} \right \|}{\sqrt{s}} 
	\end{equation}
	where  $C_1$ only depends on $\delta_{2s}$ and $\bm c^*$ is the vector $\bar{\bm c}$ with all but the s-largest entries set to be zero. If $\bar{\bm c}$ is s-sparse, then the recovery is exact.  
\end{theorem}
Since calculating the isometry constant is an NP-complete problem, one can use the following  theorem which  provides a probabilistic upper bound on the isometry constant for bounded orthonormal systems. An orthonormal system  $\{ \psi_{n}\}_{1\leqslant n \leqslant K} $ with respect to density $\rho (\xi)$ is a bounded orthonormal system if it satisfies
\begin{equation} \label{boundedorthogonalgeneral}
\underset{ 1 \leqslant n \leqslant K}\sup \left \| \psi_{n} \right \|_{\infty}= \ \underset{ 1 \leqslant n \leqslant K}\sup  \ \ \underset{ \bm \xi \in \textrm{supp}(\rho)}\sup
\left | \psi_{n}( \xi) \right | \leqslant L.
\end{equation}

\begin{theorem}[\cite{rauhut2010compressive}]\label{thm.coherence}
	Let $\{\psi_{n}\}_{1\leqslant n \leqslant K}$ be a bounded orthonormal system. Also let  $\bm \Psi \in \mathbb{R}^{M \times K}$ be a measurement matrix with entries $\left \{\psi_{ij}=\psi_{j}( \xi ^{(i)})\right \}_{1\leqslant i\leqslant M, 1\leqslant j\leqslant K}$, where $\xi ^{(1)}, ..., \xi ^{(M)}$ are random samples drawn from measure $\rho$. Assuming that 
	\begin{equation}
	M\geqslant C\delta^{-2}L^2s\log^3(s)\log(K),
	\end{equation}
	then with probability at least $1-K^{\beta \log^3(s)}$, the isometry constant $\delta_s$ of $\frac{1}{\sqrt{M}}\bm\Psi$ satisfies $\delta_s \leqslant \delta$.  $C, \beta > 0 $ are universal constants. 
\end{theorem}

  It should be noted that $L$ is the smallest such constant for which inequality \ref{boundedorthogonalgeneral} holds. Hereinafter, let us refer to  bound $L$ as the  local-coherence.

\subsection{Enhanced sampling strategies}
Theorem \ref{thm.coherence} has motivated new sampling approaches, including coherence-optimal sampling introduced in \cite{hampton2015compressive}. In this approach, instead of directly sampling from the probability measure, $\rho(\bm \xi)$, samples are drawn from a different optimal probability measure, $\rho_{\text{o}}(\bm \xi)$,  which was shown to result in the lowest local-coherence.  The optimal measure $\rho_{\text{o}}(\bm \xi)$ is constructed according to
\begin{equation}
	\rho_{\text{o}}(\bm \xi)= C^2 \rho(\bm \xi) B^2(\bm \xi),
\end{equation}
where $C$ is a normalizing constant and 
\begin{equation}
	B(\bm \xi):= \underset{\left \| \bm \alpha \right \|_1 \leqslant k}{\textrm {max}}\left |  \psi_{\bm \alpha}(\bm \xi)\right |. 
\end{equation}
Corresponding to this new probability measure, a weight function should be used to maintain asymptotic  orthogonality between the polynomial bases. This weight function is given by
\begin{equation} \label{weight}
	w(\bm \xi)=\frac{1}{B(\bm \xi)}. 
\end{equation}
Accordingly, the following weighted $\ell_1$-minimization is  solved for sparse recovery
\begin{equation} \label{eq:weightedl1min}
	\underset{\bm c}{\textrm{min}} \left \| \bm c \right \|_{1} \quad \textrm{subject to} \quad  \left \|\bm W \bm \Psi \bm c- \bm W \bm u \right \|_{2} \leq \epsilon,
\end{equation}
where $\bm W$ is the $M \times M$ diagonal weight matrix, with $\bm W(i,i)= w(\bm \xi^{(i)})$ for $i=1,\cdots, M$. 

It has been shown that coherence-optimal sampling can significantly improve the accuracy of sparse recovery for PCE \cite{hampton2015compressive}. In \cite{alemazkoor2017near}, a near-optimal sampling approach was proposed, which aimed to simultaneously improve the local-coherence and the cross-correlation properties of the resulting measurement matrix. In this approach,  first a large pool of sample candidates are drawn from the coherence-optimal sampling distribution, out of which the final sample set of size $M$ is selected such that the cross-correlation properties of the resulting measurement matrix are improved. Numerical examples have shown that near-optimal sampling can significantly improve the accuracy of sparse PCE recovery for system responses with various combinations of dimensionality and  order. 

In this work, we introduce an alternative preconditioning approach for improving the properties of measurement matrix. This approach can particularly address the case where samples have already been drawn using expensive simulations or costly experiments and sampling strategies have become inapplicable. Next section introduces our proposed design approach for  the  preconditioning matrix.   

\section{A preconditioning scheme } \label{sec:preconditioning}
In Sections \ref{recovery}, we discussed the relevant properties of the measurement matrix $\bm \Psi$ that can impact the recovery accuracy, and discussed how one can  enhance the recovery accuracy  using advanced sampling strategies \cite{hampton2015compressive, tang2014subsampled, alemazkoor2017near}. However, when the sample points are already obtained, the only way to improve the properties of $\bm \Psi$  is through preconditioning. To this end, one can multiply the measurement matrix with a preconditioning matrix, $\bm P \in \mathbb{R}^{M \times M}$,  such that the preconditioned measurement matrix $\bm P \bm \Psi$  is less coherent than the original measurement matrix. One can then solve the following $\ell_1$-minimization
\begin{equation}
\label{eq:Preconl1min}
\underset{\bm c}{\textrm{min}} \left \| \bm c \right \|_{1} \quad \textrm{subject to} \quad  \left \|\bm P \bm \Psi \bm c- \bm P \bm u\right \|_{1} \leq \epsilon. 
\end{equation}
The weight matrix in (\ref{eq:weightedl1min}) can be thought of as a preconditioning matrix, which improves the properties of the measurement matrix. However, by definition it is a diagonal matrix, with its diagonal terms $w(i,i)$ being a function of the $i$th sample location. Here, we do not force $\bm P$ to be diagonal, neither do we constrain it to be a function of sample locations. In what follows, we discuss the relevant criteria used in our design of preconditioning matrix. 

Let us first define $\bm D \triangleq \bm P \bm \Psi  \in \mathbb{R}^{M \times K}$ and call it the \textit{equivalent measurement matrix}.  One of the main properties of $\bm D$ that directly impacts the recovery accuracy is its spark. The spark of a matrix is the smallest number of its columns that are linearly dependent. It is known that $\ell_1$ minimization is guaranteed to recover an $s$-sparse signal vector if spark($\bm D$)$>2s$. Designing the preconditioning matrix such that the spark of the equivalent measurement matrix is maximized would allow the exact recovery of a larger set of signals. However, computing the spark of a matrix is an NP-hard problem. Alternatively, one can analyze recovery guarantees using other properties which are easier to compute. One such property is the mutual-coherence, which for a given matrix is defined as the maximum absolute normalized inner product, i.e. the cross-correlation, between its columns \cite{doostan2011non, donoho2006stable}.  Let $D_{1}, D_{2}, ..., D_{K} \in \mathbb{R}^{M}$ be the columns of matrix $\bm D$. The mutual-coherence of matrix $\bm D$, denoted by $\mu(\bm D)$, is then given by
\begin{equation} \label{eq:coherence}
\mu(\bm D):= \underset{1\leqslant  i, j\leqslant  K, i\neq j}{\max}\frac{|D_{j}^{T}D_{i}|}{\begin{Vmatrix}D_{j}\end{Vmatrix}_{2}\begin{Vmatrix}D_{i}\end{Vmatrix}_{2}}.
\end{equation}\\ 
Mutual-coherence gives a lower bound for spark as follows \cite{elad2010sparse}
\begin{equation} \label{spark}
	\text{spark}(\bm D)\geqslant 1+\frac{1}{\mu(\bm D)}.
\end{equation}
 It may be concluded that preconditioning matrix $\bm P$ should be designed in a way that mutual-coherence of $\bm D$ is minimized. However, it has been observed that minimizing mutual-coherence does not necessarily improve the recovery accuracy of compressive sampling \cite{li2013projection,elad2007optimized}. This is because mutual coherence only serves as a lower bound on the spark, and minimizing mutual-coherence is   optimizing only for the worst-case scenario and fails to reflect upon other possibilities for improving compressive sampling performance  \cite{elad2007optimized}.  
 
 Theorem \ref{thm.coherence} suggests that optimizing the preconditioning matrix so that isometry constant of the equivalent measurement matrix is minimized will be effective.  However, as previously mentioned, calculating isometry constant for a given matrix is an NP-complete problem. If one considers all the column-submatrices of the equivalent measurement matrix that has  $s^{*} \leqslant s$ columns, it is known that the eigenvalues of these column-submatrices are bounded by the isometry constant. Therefore, one may suggest to minimize the condition number of all column-submatrices with $s$ or fewer columns. However, calculating such combinatorial measure is not a trivial task either and can be computationally impossible \cite{elad2007optimized}. 
 
 As a result, establishing a single matrix property that (i) is directly associated with the accuracy of the compressive sampling method and (ii) is easily computable still remains an open challenge  \cite{li2013projection,elad2007optimized}. To sidestep this challenge, efforts have focused on identifying properties or measures that are relatively better than the mutual-coherence. For example, the average coherence of the equivalent measurement matrix can be minimized \cite{elad2007optimized}. Alternatively, one can minimize the distance between the Gram matrix, $\bm D^T \bm D$, and the identity matrix or an equiangular tight frame (ETF) Gram matrix \cite{li2013projection, abolghasemi2012gradient, zelnik2011sensing,tian2016orthogonal}. ETF matrices are matrices consisting of unit vectors whose maximum cross-correlation achieves the smallest possible value, and as such are ideal measurement matrices for sparse recovery.
 
 When  $\bm \Psi \bm c+\bm e= \bm u$, where $\bm e$ is the vector representing noise and/or modeling error, a preconditioning matrix obtained by minimizing the distance between Gram matrix and ETF Gram matrix may magnify the noise while reducing the magnitude of observation vector. In other words, it's possible that $\left \| \bm P \bm e \right \|_2$ is large while $\left \| \bm P \bm u \right \|_2$ is relatively small when $\bm P$ is not properly chosen, thereby deteriorating the signal-to-noise ratio and in turn the PCE approximation accuracy.  
 
 Similar issue exists for designing a projection (or sensing) matrix in signal processing application. Recent studies have proposed to design a projection matrix such that it improves incoherence properties of equivalent dictionary, and its multiplication by the noise vector results in a vector with small magnitude \cite{hong2016efficient,li2015designing,hong2018efficient}. However, it's more challenging to design a preconditioning matrix for an underdetermined regression problem as both sides of the equation, $\bm \Psi \bm c+\bm e= \bm u$, are multiplied by the preconditioning matrix. Therefore, designing a preconditioner such that $\left \| \bm P \bm e \right \|_2$ is minimized may result in undesirably small magnitude for $\left \| \bm P \bm u \right \|_2$. 
 
Alternatively, one may propose designing the  preconditioning matrix such that $\left \| \bm P \bm u \right \|_2$ is maximized. However, this doesn't provide control over how much the noise vector gets magnified.   To prevent undesirable relative magnification for $\left \| \bm P \bm e \right \|_2$ and $\left \| \bm P \bm u \right \|_2$, we seek to design a preconditioning matrix  that  (1) results in a small distance between the Gram matrix and the ETF Gram matrix and (2) is not too much different from the identity matrix. We choose identity matrix since it is guaranteed not to disproportionally magnify the noise vector as compared with the measurement vector. The choice of identity matrix as a reference matrix is robust in the sense that it does not assume any particular choice for the measurement noise distribution. Following these reasonings, we form the following problem for the design of preconditioning matrix:
 \color{black}
\begin{equation}\label{eq:premin}
\min_{\bm P \in \mathbb{R}^{M \times M}, \bm G \in H_{\mu_E}} f(\bm G,\bm P)=\left \|\bm G - \bm \Psi^T \bm P^T \bm P \bm \Psi \right \|_{F}+ \lambda \left \|\bm I- \bm P\right \|_{F},
\end{equation}
where $\left \|.\right \|_{F}$ denotes the Frobenius norm and $\bm G$ is the targeted ETF Gram matrix which belongs to the space
\begin{equation}
H_{\mu_E}=\left \{ \bm G \in \mathbb{R}^{K \times K}: \bm G= \bm G^T,  G_{ii}=1, \underset{i\neq j }{\text{max}} \left | G_{ij} \right|< \mu_E \right \},
\end{equation}
where $\mu_E$ is the lower bound for the mutual-coherence of matrix $\bm \Psi$ of dimension $M \times K$ given by
\begin{equation}
\mu_E := \sqrt{\frac{K-M}{M(K-1)}}.
\end{equation}
 When $\lambda=0$, the problem defined in (\ref{eq:premin}) is usually solved using alternating minimization method, where $\bm P$ is updated using a gradient-based algorithm \cite{abolghasemi2012gradient,hong2016efficient}. Here, we use the similar approach to solve \ref{eq:premin} with $\lambda>0$. The gradient of $f(\bm G, \bm P)$ with respect to $\bm P$ is given by
 \begin{equation}\label{eq:derv}
 \frac{\partial f}{\partial \bm P}= \frac{\partial}{\partial \bm P}\text{Tr}\left \{ (\bm G - \bm \Psi^T \bm P^T \bm P \bm \Psi)^T (\bm G - \bm \Psi^T \bm P^T \bm P \bm \Psi)\right \} + \lambda \frac{\partial}{\partial \bm P}\text{Tr}\left \{ (\bm I- \bm P)^T(\bm I- \bm P)\right \},
 \end{equation}
where $\text{Tr}\left \{  \cdot\right \}$ denotes the  trace operator. Using matrix derivative rules, Equation $\ref{eq:derv}$ can be further simplified to
\begin{equation}\label{eq:dervSim}
\frac{\partial f}{\partial \bm P}= 4\bm P \bm \Psi(\bm \Psi^T \bm P^T \bm P \bm \Psi - \bm G)\bm \Psi^T+2 \lambda(\bm P-\bm I).
\end{equation}
Given the gradient of $f(\bm G,\bm P)$, the conjugate gradient method is used to solve \ref{eq:premin} for a fixed $\bm G$. We then use an alternating approach to solve (\ref{eq:premin}). Our alternating approach keeps one of the two variables in  \ref{eq:premin} constant and solves the problem by varying the other parameter. When $\bm G$ is fixed, as we explained, the conjugate gradient method  is used to evaluate the solution of \ref{eq:premin}. Once $\bm P$ is fixed, the optimal $\bm G$ can be simply obtained by projecting the Gram matrix of column-normalized equivalent measurement matrix, $\tilde{\bm D}$, onto the set $H_{\mu_E}$ \cite{abolghasemi2012gradient} according to:
\begin{equation}
 G_{ij}= \left\{\begin{matrix}
1, & i=j,\\ 
\tilde{G}_{ij},& i\neq j, |\tilde{ G}_{ij} |\leq\mu_E,\\ 
 \text{sign}(\tilde{ G}_{ij}).\mu_E, & i\neq j, | \tilde{ G}_{ij} |> \mu_E, 
\end{matrix}\right.
\end{equation}
where sign($\cdot$) is a sign function, and $\tilde{\bm G}=\tilde{\bm D}^T \tilde{\bm D}$. Algorithm \ref{pseudocode} shows  the alternating approach to solve \ref{eq:premin}. 
 \begin{algorithm}[H] 
 	\caption{The alternating solution for minimization problem \ref{eq:premin}}\label{pseudocode}
 	\begin{algorithmic}[1]
 		\State Initialization: Set a convergence threshold $\delta_{\text{thr}}$; set $\bm P_0$ to be a random matrix, and $\bm G_0$ to be the identity matrix.
 		
 		\While {$\delta_t<\delta_{\text{thr}}$} 
 		\State  Form  equivalent measurement matrix: $\bm D=\bm P_{t-1}  \bm \Psi $.
 		\State Form column-normalized equivalent measurement matrix:  $\tilde{\bm D}$.
 		\State $\tilde{\bm G}=\tilde{\bm D}^T \tilde{\bm D}$.
 		\State $
 		G_t(i,j)= \left\{\begin{matrix}
 		1, & i=j,\\ 
 		\tilde{G}_{ij},& i\neq j, |\tilde{ G}_{ij}|\leq\mu_E,\\ 
 		\text{sign}(\tilde{ G}_{ij}).\mu_E, & i\neq j, |\tilde{ G}_{ij} |> \mu_E.
 		\end{matrix}\right.$
 		\State Solve $\bm P_t=  \underset{\bm P \in \mathbb{R}^{M \times M}}{\text{argmin}} f(\bm G_t, \bm P )$ using conjugate gradient method.
 		\State  $\delta_t=[{f(\bm G_t,\bm P_t)-f(\bm G_{t-1},\bm P_{t-1})}]/{f(\bm G_{t-1},\bm P_{t-1})}.$
 		
 		\EndWhile
 		
 	\end{algorithmic}
 \end{algorithm} 
 Algorithm \ref{pseudocode} is used to solve \ref{eq:premin} for a fixed $\lambda$ value. We propose a cross-validation approach that identifies the  $\lambda$ value that results in best PC  approximation accuracy.  To do so we first randomly divide the $M$ samples into $M_{\text{tr}}$ training samples and $M_{\text{val}}$ validation samples. Then, for each $\lambda$ value, using all the $M$ samples, we design a preconditioning matrix which leads to an equivalent measurement matrix $\bm P \bm \Psi$. We then select $M_{\text{tr}}$ rows of this measurement matrix, based on the training samples, to form the training submatrix of size $M_{\text{tr}} \times K$.   Using this training measurement submatrix we estimate the expansion coefficients  and calculate the validation error over the validation samples. Once this is done for all the choices of $\lambda$ values, we then choose the $\lambda$ value and its associated preconditioning matrix that results in the smallest validation error. Algorithm \ref{pseudocode_crossvalidation} shows the steps for the proposed cross-validation algorithm. In this pseudo-code,  superscripts `tr' and `val' denote the row-submatrices associated with training and validation samples, respectively. 
 \begin{algorithm}[H] 
 	\caption{Cross-validation algorithm for weigh parameter $\lambda$}\label{pseudocode_crossvalidation}
 	\begin{algorithmic}[1]
 		\State Initialization: Form a set of $N$ candidate values for $\lambda$: $\left \{\lambda_1,\cdots,\lambda_N\right \}.$ 
 		\State Randomly divide the available $M$ samples into $M_{\text{tr}}$ training samples and $M_{\text{val}}$ validation samples.
 		\For {$i=1:N$} 
 		\State Solve $\bm P_i = \underset{\bm P \in \mathbb{R}^{M \times M}, \bm G \in H_{\mu_E}}{\text{argmin}}\left \|\bm G - \bm \Psi^T \bm P^T \bm P \bm \Psi \right \|_{F}+ \lambda_i \left \|\bm I- \bm P\right \|_{F}$.
 		\Comment{Use Algorithm \ref{pseudocode}}
 		\State Precondition the problem: $\bm D_i=\bm P_i \bm \Psi$ and $\bm u_{P_i}=\bm P_i \bm u$.
		\State Form  measurement submatrix $\bm D_i^{\text{tr}}$ and preconditioned measurement subset $\bm u_{P_i}^{\text{tr}}$ at  training samples.
 		\State Solve $\bm c_i= \underset{\bm c}{\textrm{argmin}} \left \| \bm c \right \|_{1} \quad \textrm{s.t.} \quad  \left \|\bm D_i^{\text{tr}} \bm c- \bm u_{P_i}^{\text{tr}}\right \|_{2} \leq \epsilon_i$, \Comment{Choose $\epsilon_i$ using separate cross-validation}
 		\State Calculate cross-validation error $e^{CV}_i = \left \|\bm \Psi^{\text{val}} \bm c_i- \bm u^{\text{val}}\right \|_{2}$.
 		\EndFor
 		\State Choose $\lambda_i$ that results in the smallest cross-validation error $e^{CV}_i$.
 	\end{algorithmic}
 \end{algorithm}
When it is not possible to select a preconditioning matrix without undesirably magnifying the noise, Algorithm \ref{pseudocode_crossvalidation} will identify a large value for $\lambda$, thereby selecting a preconditioning matrix close to an identity matrix. Therefore, it can be argued that the proposed preconditioning approach is expected to either improve or in the worst case preserve the approximation accuracy. In what follows, using numerical examples, we demonstrate the potential accuracy improvement that can be achieved by proposed preconditioning algorithm.
 
\section{Numerical illustration} \label{sec:Numerical examples}
To demonstrate the improvement in approximation accuracy using the proposed preconditioning scheme, we first conduct a rather comprehensive numerical study and report phase transition diagrams for Hermite- and Legendre-based target functions with different combinations of polynomial dimension, $d$, and polynomial order, $k$. As additional numerical examples, we also consider three different types of target functions: (i)  functions that are exactly sparse with respect to PC bases together with noiseless measurements, (ii) functions that are approximately sparse with respect to PC bases together with noiseless measurements, and (iii) functions that are exactly sparse  with respect to PC bases together with noisy measurements.  

In all the examples, we column-normalize the measurement matrix to prevent $\ell_1$ minimization biasing towards  columns with large norms. Optimization problems (\ref{eq:l1minNo}) and (\ref{eq:Preconl1min}) are solved using the SPGL1 package \cite{van2007spgl1}.  We use the default setting of SPGL1 package, except for the maximum number of iteration, tolerance for identifying a basis pursuit solution, \verb bpTol ,  and optimality tolerance, \verb optTol . We set maximum number of iterations, \verb bpTol , and  \verb optTol , respectively equal to $50 M$ (with $M$ being the number of samples), $10^{-8}$, and $10^{-6}$. The conjugate gradient-based  \textit{MinFunc} toolbox \cite{schmidt2006minfunc} is used to solve (\ref{eq:premin}). In Algorithm \ref{pseudocode}, we set the convergence threshold to be $10^{-2}$. Finally, for the cross validation algorithm for $\lambda$ value, three fourths of the samples are used for training and the rest  for validation. 
\subsection{Phase transition diagrams}  
As the first illustration case, consider the target function to be in fact a polynomial expansion, i.e. $f(\bm \Xi)= \sum_{j=1}^{K} \bm c_{\bm \alpha^j}\bm\Psi_{\bm \alpha^j}(\bm \Xi)$, where the known coefficient vector $\bm c$ is exactly $s$-sparse. Our objective is to evaluate recoverability of  $\bm c$ using only $M$ samples. To perform a comprehensive recoverability evaluation,  we consider several scenarios with varying sparsity and sample size. Specifically, let us consider a unit square $[0, 1]^2$, where the x-axis shows  undersampling rates, $M/K$, and the y-axis shows sparsity rates, $s/M$. This unit square, which constitutes the domain for phase transition diagrams, typically partitions into three regions: (1) a region where the probability of accurate recovery is near one; (2) a region where the probability of accurate recovery is near zero; (3) a narrow transition region \cite{donoho2005neighborliness, donoho2010precise}.  Furthermore, we consider scenarios where order and dimension of polynomial expansion are varied. For each choice of dimension and order, we divide the unit square to a $50 \times 50$ grid. At each point in the grid, i.e. for each combination of $d, k, s$ and $M$,  we study the success rate of the   $\ell_1$-minimization on 100 ``trial" target expansions. Each trial target expansion is created by randomly selecting   $s$  coefficients and assigning them  to values drawn from a standard normal distribution, while setting the remaining coefficients equal to zero. For each trial target expansion, the recovery is determined to be successful if ${\left \| \bm c- \bar{\bm c}\right \|_2}/{\left \| \bar{\bm c} \right \|_2} < 10^{-3} $, where $\bar{\bm c}$ is the coefficient of the target expansion and $\bm c$ is the  solution of the $\ell_1$-minimization. The recovery probability is then defined as the ratio of  successful recoveries among all  recoveries for the 100 trial target expansions.  Finally, the transition diagrams are created by connecting the points at which the successful recovery probability is estimated to be 50\%.  In this work, we  compare phase transitions diagrams for standard and coherence-optimal sampling with and without preconditioning. It should be noted that in generating  `coherence-optimal' results  (obtained without our preconditioning scheme) we still apply the weight matrix, $\bm W$, as in (\ref{weight}) and solve (\ref{eq:weightedl1min}). This weight matrix may also be referred to as a preconditioning matrix. However, in this paper,  only the results labeled   'preconditioned'    have been obtained from our proposed preconditioning approach shown in Algorithm \ref{pseudocode_crossvalidation}.

\begin{figure} [H]
  	
  	\begin{subfigure}[t]{0.33 \linewidth}
  		\includegraphics[width=1\linewidth]{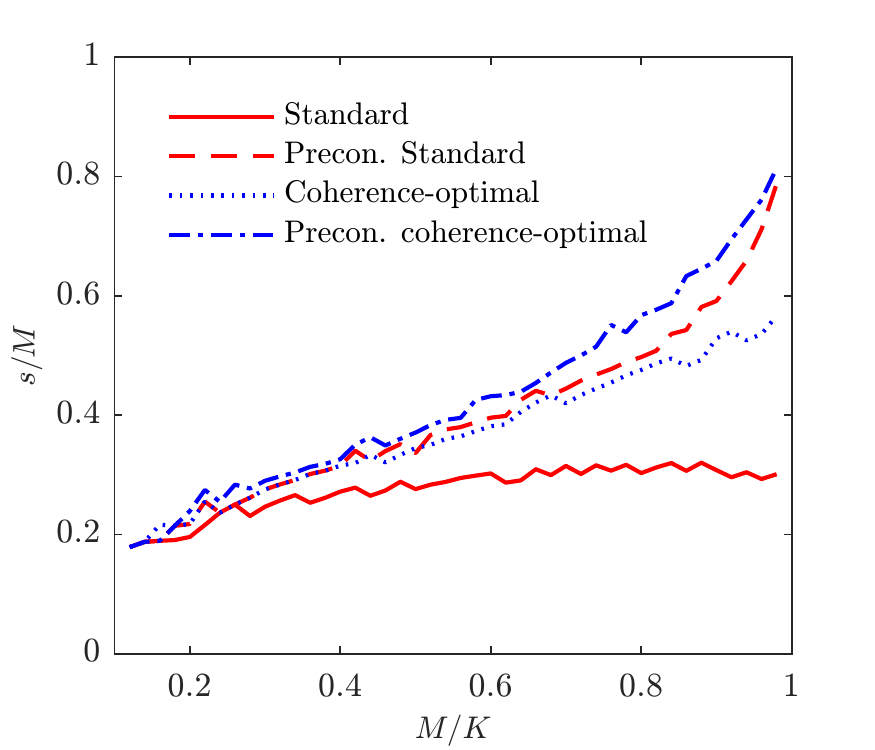}
  		\caption{$d=2, k=20$}
  		\label{fig:phased2p20}		
  	\end{subfigure}
  	\quad
  	\begin{subfigure}[t]{0.33 \linewidth}
  		\includegraphics[width=1\linewidth]{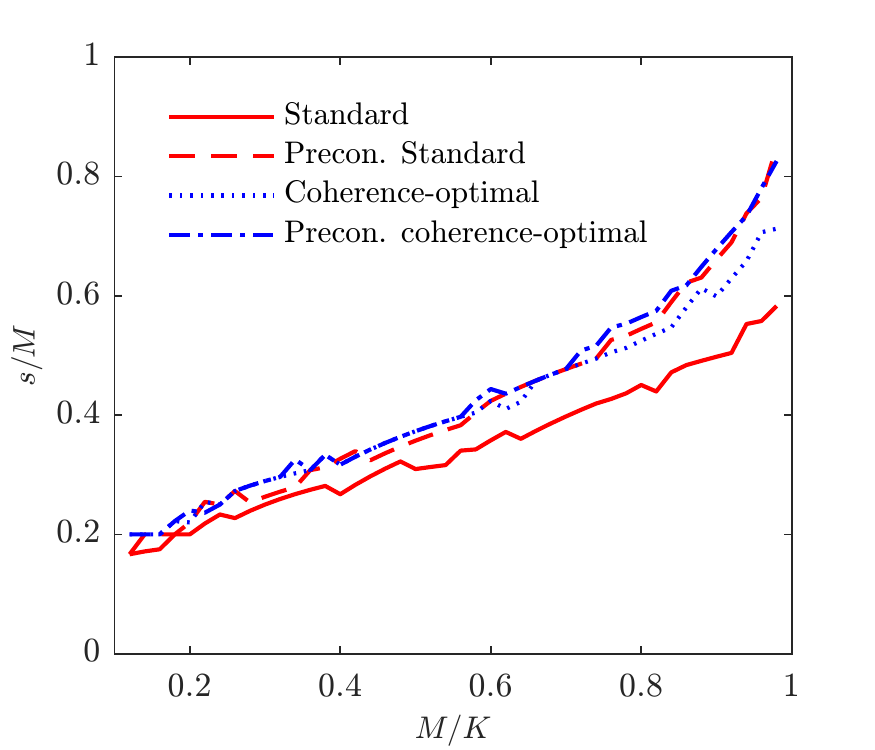}
  		\caption{ $d=5, k=5$}
  		\label{fig:phased5p5}
  	\end{subfigure} 
  	\quad
  	  	\begin{subfigure}[t]{0.33 \linewidth}
  	  		\includegraphics[width=1\linewidth]{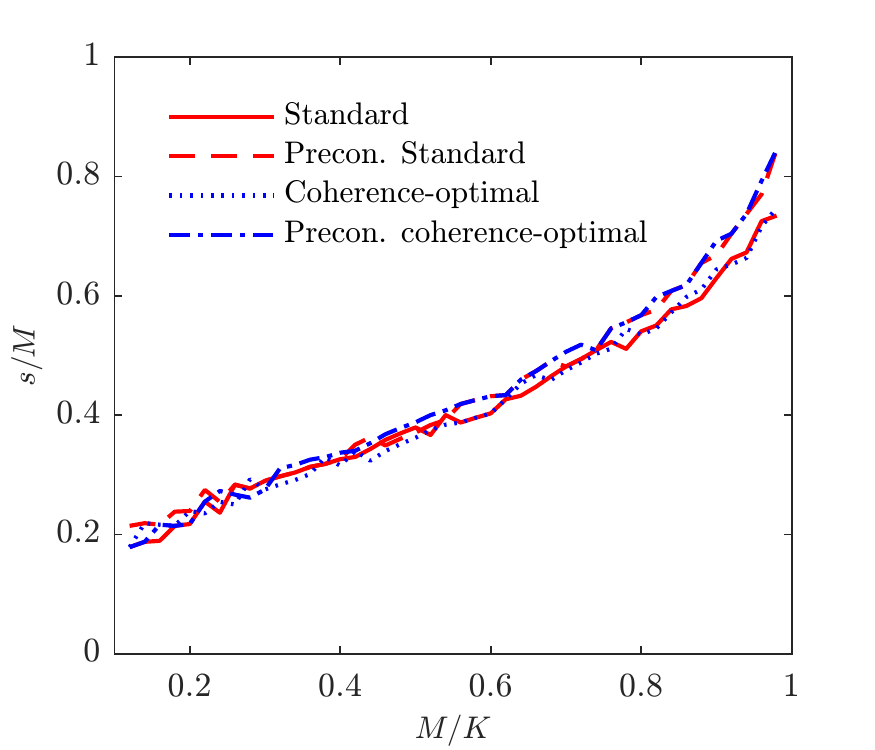}
  	  		\caption{$d=20, k=2$}
  	  		\label{fig:phased20p2}
  	  	\end{subfigure}
  	
  	\caption{ Comparison of  phase transition diagrams for standard and coherence-optimal sampling with and without preconditioning for sparse recovery of Legendre-based PCE with three combinations of dimension $d$ and order $k$.}
  	\label{fig.phase}
  \end{figure}

Figure \ref{fig.phase} includes the phase transition diagrams for 3  combinations of dimension and order for Legendre polynomial expansions and shows that preconditioning significantly improves the recovery accuracy. In cases where $d \ll  k $ or $k \approx d$ preconditioned coherence-optimal sampling outperforms preconditioned standard sampling. This is because  coherence-optimal sampling is expected to outperform standard sampling more significantly in these cases. For large values of $k$ coherence-optimal sampling distribution coverages to Chebyshev distribution, which is the optimal sampling distribution as $k$ goes to infinity \cite{jakeman2017generalized}. As the dimensionality increases and the polynomial order decreases  the coherence-optimal sampling distribution becomes similar to uniform distribution. This can be explained by referring  to  the coherence-optimal distribution given by $\rho_{\text{o}}(\bm \xi)= C^2 \rho(\bm \xi) B^2(\bm \xi)$,   where $C$ is a normalizing constant and $B(\bm \xi)$ is the largest absolute value of all the polynomial basis functions. For Legendre-based expansions,   $\rho_{\text{o}}(\bm \xi) \propto B^2(\bm \xi$). For $d \ll  k $ or $d \approx k $, this distribution is expected to peak when  $\bm \xi$ values in all dimensions are close to the boundary. To explain this, we note that one-dimensional $\alpha_i$-degree Legendre polynomials reach their maximum absolute values at their support boundaries, and that $B(\xi_i) \leqslant (2 \alpha_i+1)^{\frac{1}{2}}$  \cite{yan2012stochastic}. Therefore, for   $d$-dimensional Legendre polynomials with the multi-index $\bm \alpha$, we have
\begin{equation}\label{eq.legendrebound}
B(\bm \xi) \leqslant \prod_{i=1}^{d} (2 \alpha_i+1)^{\frac{1}{2}}.
\end{equation}

When $d \ll k$, the expansion will include full-dimensional bases (where all the dimensions are present) and therefore $B(\bm \xi)$ is peaked when $\bm \xi$ values in all dimensions are close to the boundary. However, when  $d \gg k$, the expansion will not include full-dimensional bases; it will have at most $k$-dimensional bases, and the full-dimensional bound of Equation~\ref{eq.legendrebound} will never be triggered. As a result, $B(\bm \xi)$ is peaked at samples of $\bm \xi$ that have $k$ entries  close to the boundary, and the remaining $d-k$ entries can be randomly located anywhere in the parameter space. Consequently, for large values of $d$ and small values of $k$, $\rho_{\text{o}}(\bm \xi)$ becomes similar to uniform distribution  and coherence-optimal sampling produces results similar to those obtained by standard sampling.

\begin{figure} [H]
    	
    	\begin{subfigure}[t]{0.33 \linewidth}
    		\includegraphics[width=1\linewidth]{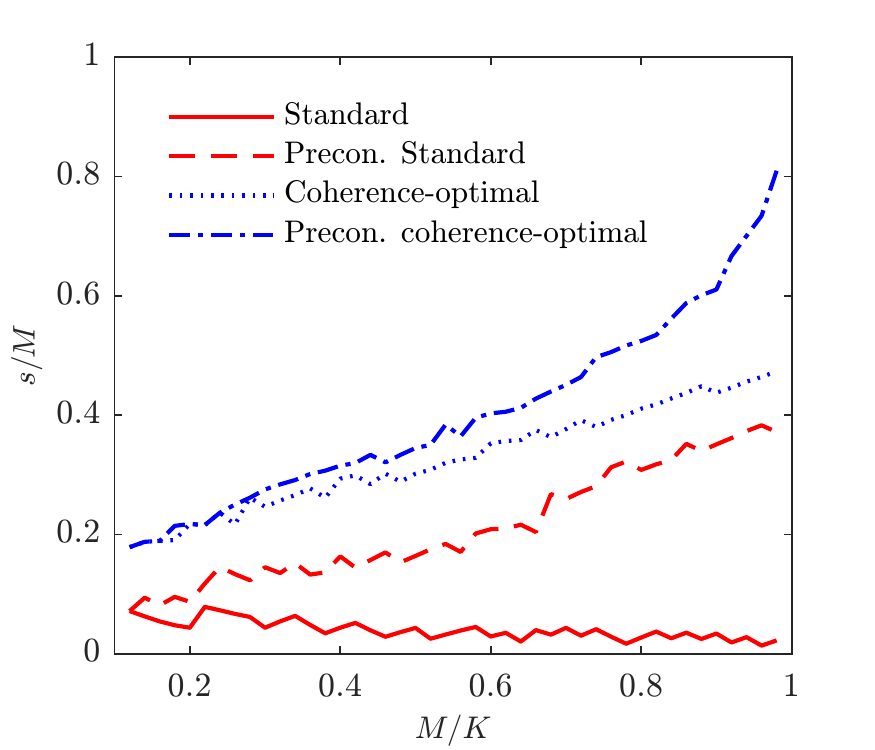}
    		\caption{ $d=2, k=20$}
    		\label{fig:phased2p20.her}		
    	\end{subfigure}
    	\quad
    	\begin{subfigure}[t]{0.33 \linewidth}
    		\includegraphics[width=1\linewidth]{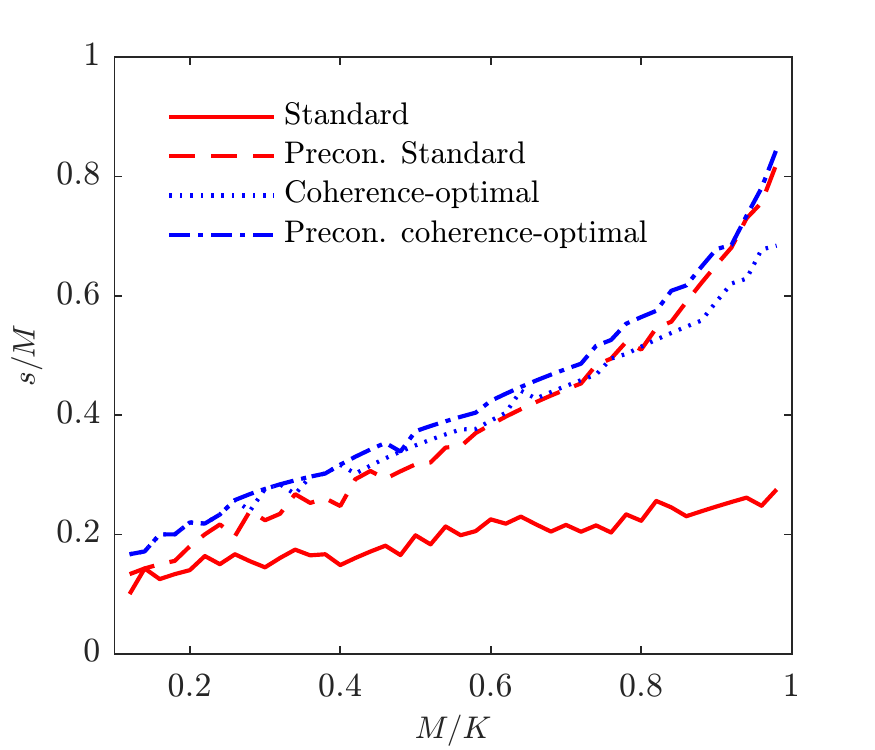}
    		\caption{$d=5, k=5$}
    		\label{fig:phased5p5.her}
    	\end{subfigure}
    	\quad
    	\begin{subfigure}[t]{0.33 \linewidth}
    		\includegraphics[width=1\linewidth]{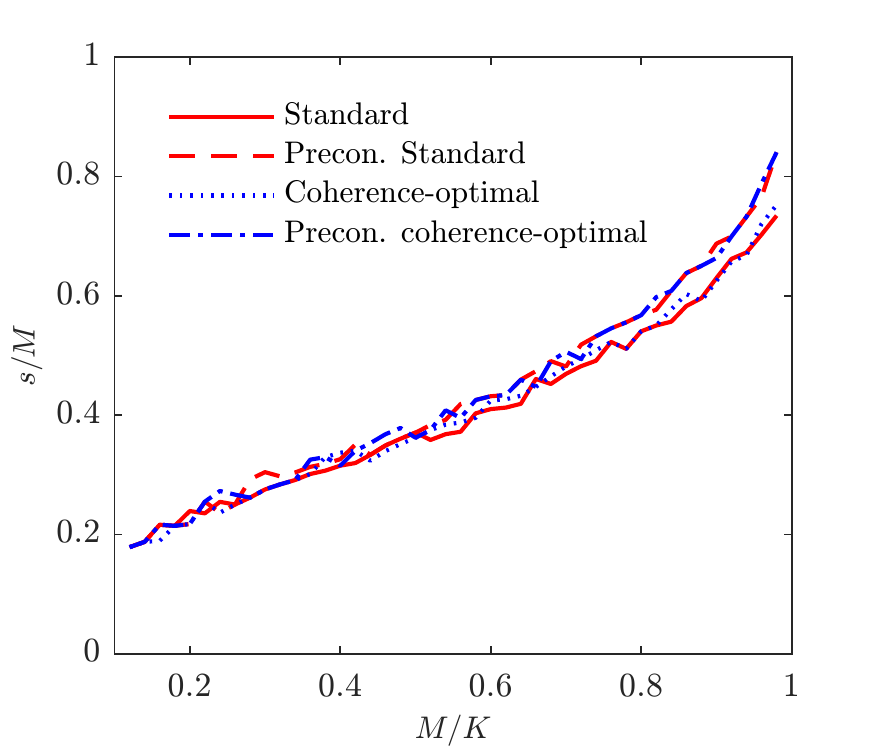}
    		\caption{$d=20, k=2$}
    		\label{fig:phased20p2.her}
    	\end{subfigure}
    	
    	\caption{ Comparison of  phase transition diagrams for standard and coherence-optimal sampling with and without preconditioning for sparse recovery of Hermite-based PCE with three combinations of dimension $d$ and order $k$. }
    	\label{fig.phase.her}
    \end{figure}

Figure \ref{fig.phase.her} includes the phase transition diagrams for Hermite polynomial functions considering the same combinations of  dimensionality and order. Again, it can be seen that advantage of coherence-optimal sampling over  standard sampling disappears  when target solutions become higher-dimensional and lower-order. Also, we observe that in all cases preconditioning significantly improves the recovery accuracy.

\subsection{Exactly sparse functions}
In this section, we consider two functions that are exactly sparse with respects to PC basis. In particular, we consider a low-dimensional high-order and a high-dimensional low-order problem.
\subsubsection{A low-dimensional high-order polynomial function} 
\begin{figure} [H]
	\begin{subfigure}[t]{0.45\linewidth}
		\includegraphics[width=1\linewidth]{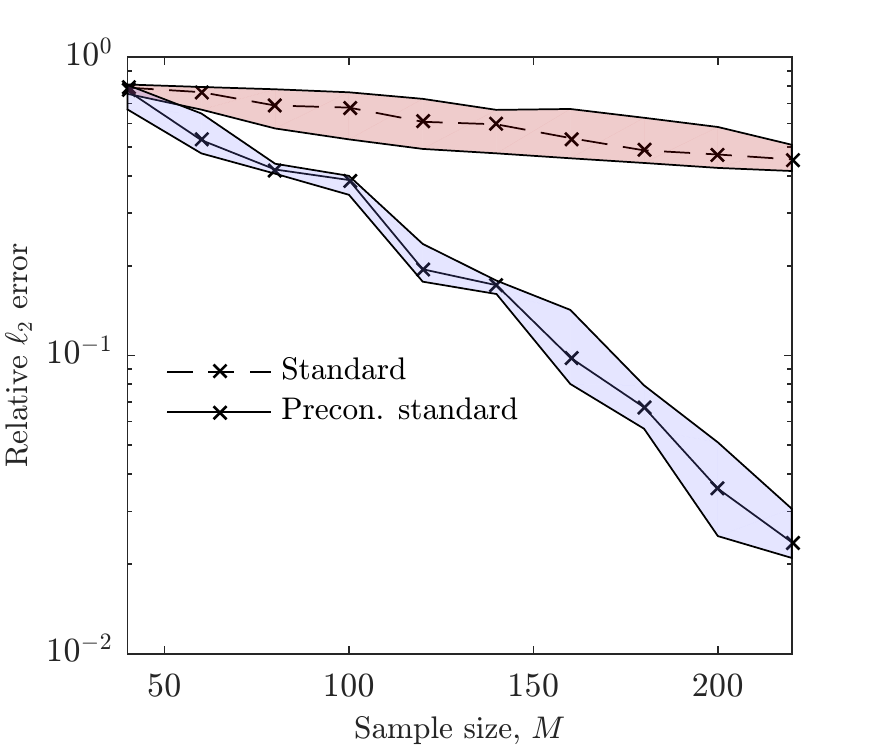}
		\caption{Standard sampling}
		\label{fig:example1-error_MC}		
	\end{subfigure}
	\begin{subfigure}[t]{0.45 \linewidth}
		\includegraphics[width=1\linewidth]{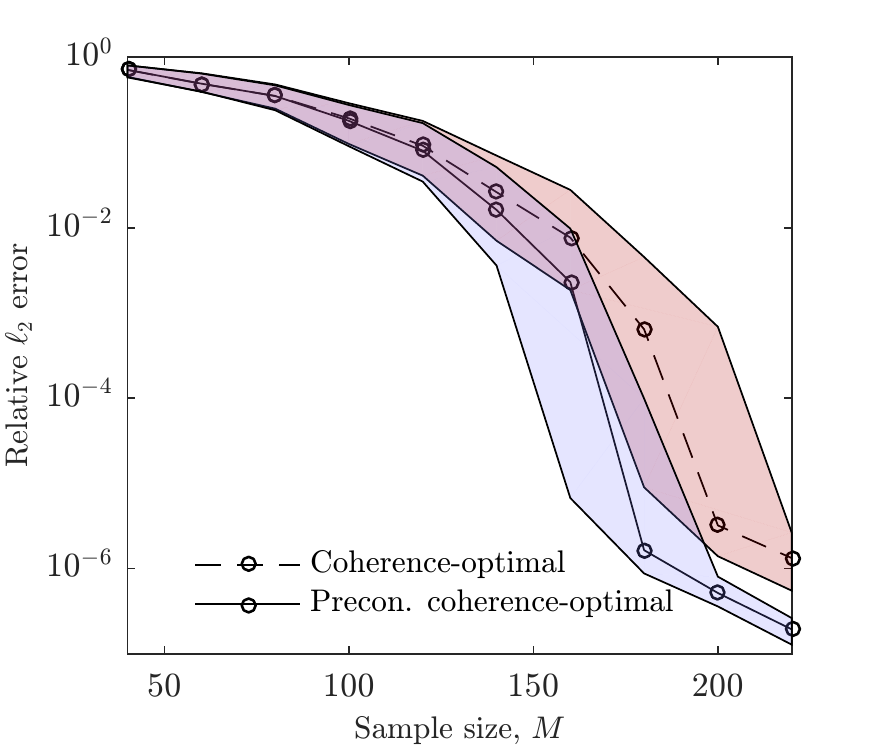}
		\caption{Coherence-optimal sampling}
		\label{fig:example1-error_CO}		
	\end{subfigure} \centering 
	\caption{Comparison of relative $\ell_2$ error with and without preconditioning for low-dimensional high-order manufactured PCE of Eq.~\ref{ex.lowDhighk} using: (a) standard sampling, (b) coherence-optimal sampling }
	\label{fig.example1}
\end{figure}

 Let $u_1(\bm \Xi)$ to be a sparse 20th-order Hermite polynomial expansion in a two-dimension random space with standard normal distribution, manufactured according to  
\begin{equation}\label{ex.lowDhighk}
	u_1(\bm \Xi)= \sum_{i=1}^{9} \Xi_1^{i}\Xi_2^{i+2}.
\end{equation}
We aim to recover the sparse vector of coefficients using a smaller set of samples than the cardinality of basis set, which is 231. We compare the approximation accuracy for standard sampling and coherence-optimal sampling with and without preconditioning. For all four approaches, we report the performance results obtained by 100 independent runs, each with an independent set of samples. Figure \ref{fig:example1-error_MC} compares the median, 1st and 3rd quantiles of relative $\ell_2$ error with and without preconditioning when standard sampling is used. Relative error is calculated as $\left \| \bm c - \bar{\bm c} \right \|_2 / \left \| \bar{\bm c} \right \|_2$, where $\bar{\bm c}$ is the exact coefficient vector and $\bm c$ is the solution of the associated $\ell_1$ minimization. Figure \ref{fig:example1-error_CO} compares the median, 1st and 3rd quantiles of relative $\ell_2$ error with and without preconditioning when coherence-optimal sampling is used. As expected, preconditioning improves the approximation accuracy for both standard and coherence-optimal sampling. Also, as it can be concluded from the phase transition diagrams in Figure \ref{fig:phased2p20.her}, the preconditioned coherence-optimal sampling results in the most accurate approximation followed by coherence-optimal and preconditioned standard sampling. It should be noted that coherence-optimal sampling significantly outperforms standard sampling for high-order Hermite polynomials. This explains the outperformance of coherence-optimal sampling over preconditioned standard sampling. Moreover, accuracy improvement achieved by preconditioned coherence-optimal sampling shows that even when samples are drawn from an optimal distribution, accuracy can still be improved using preconditioning.  
\subsubsection{A high-dimensional low-order polynomial function} 
 As a contrasting example, we consider $u_2(\bm \Xi)$ to be a sparse second-order Legendre polynomial expansion in a 20-dimensional random space with uniform density on $[-1,1]^{20}$, manufactured according to 
 \begin{equation}\label{ex.highDlowk}
 u_2(\bm \Xi)= \sum_{i=1}^{19} \Xi_i\Xi_{i+1}+\sum_{i=1}^{20} \Xi_i^2.
 \end{equation}
 
\begin{figure} [H]
	\begin{subfigure}[t]{0.45\linewidth}
		\includegraphics[width=1\linewidth]{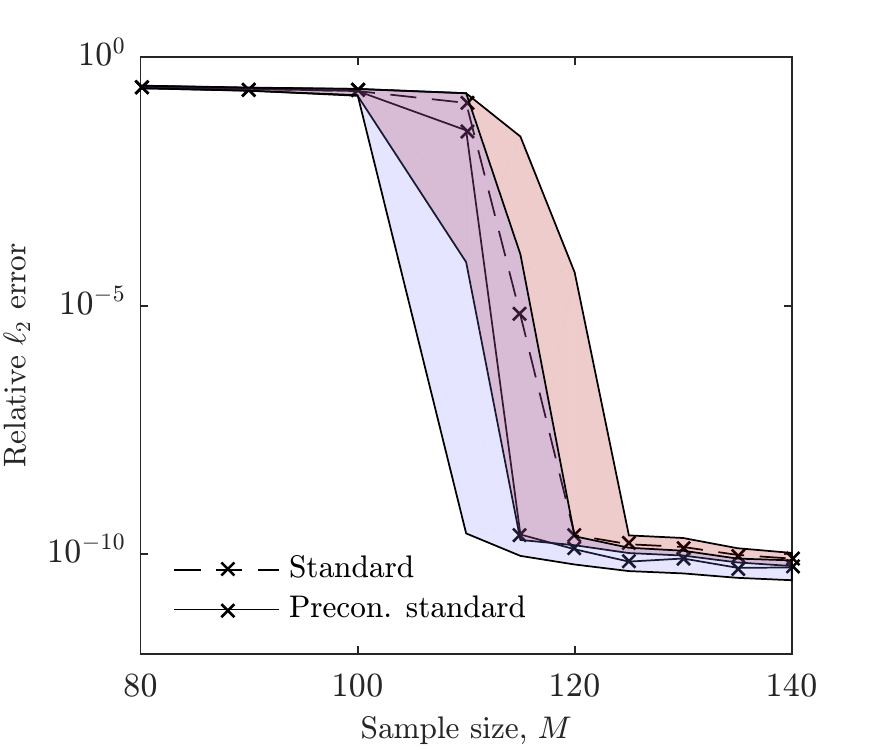}
		\caption{Standard Sampling}
		\label{fig:example2-error_MC}		
	\end{subfigure}
	\begin{subfigure}[t]{0.45 \linewidth}
		\includegraphics[width=1\linewidth]{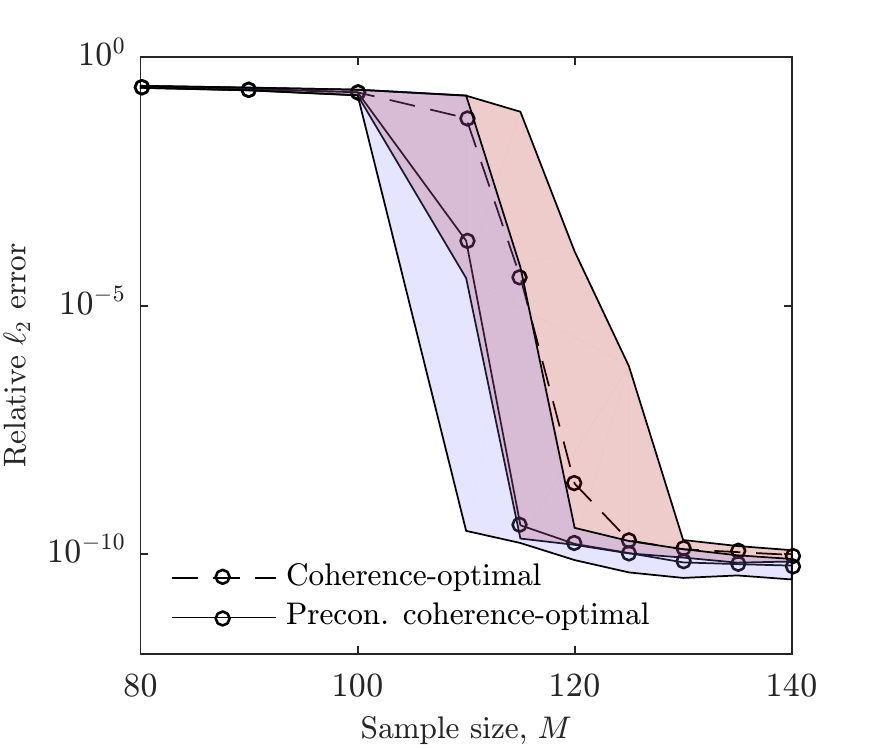}
		\caption{Coherence-optimal sampling}
		\label{fig:example2-error_CO}		
	\end{subfigure} \centering
	\caption{Comparison of relative $\ell_2$ error with and without preconditioning for high-dimensional low-order manufactured PCE of Eq.~\ref{ex.highDlowk} using: (a) standard sampling, (b) coherence-optimal sampling }
	\label{fig.example2}
\end{figure}

Similar to previous example, the cardinality of basis set is 231 and we aim to recover the vector of coefficients using a small set of samples. To compare the performance of coherence-optimal and standard sampling with and without preconditioning
on this target function, the numerical results were obtained under a setting similar to that in the previous example with 100 independent runs for all the four approaches. Figures \ref{fig:example2-error_MC} and \ref{fig:example2-error_CO} compare the median, 1st and 3rd quantiles of relative $\ell_2$ error when standard sampling and coherence-optimal sampling is used, respectively. It can be seen that the improvement in approximation accuracy achieved by coherence-optimal sampling, over standard sampling, is insignificant. However, preconditioning successfully improves the approximation accuracy.

\begin{figure} [H]
	\begin{subfigure}[t]{0.45 \linewidth}
		\includegraphics[width=1\linewidth]{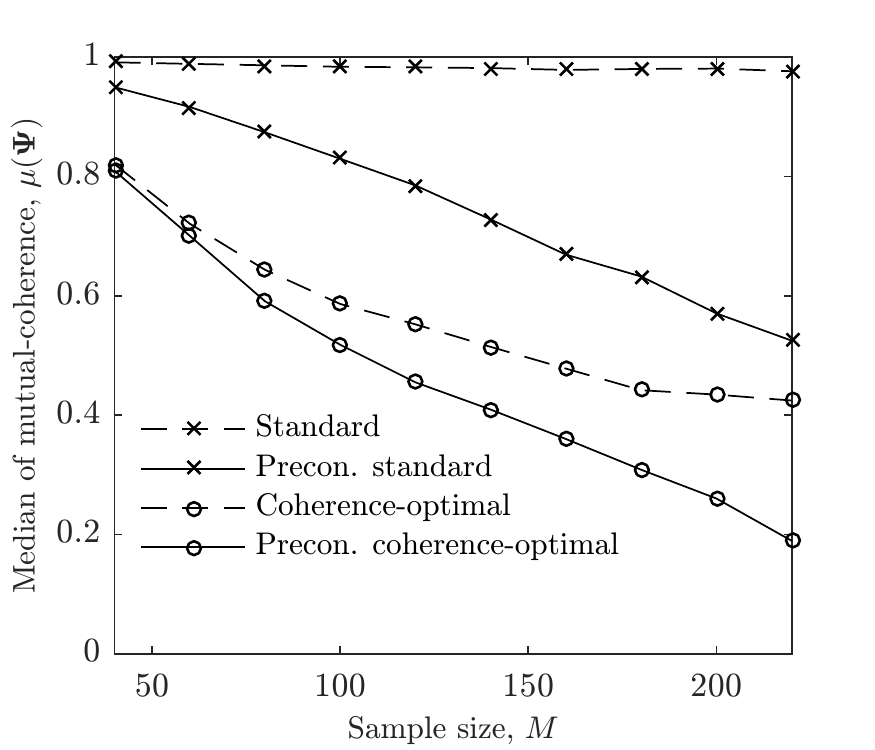}
		\caption{Low-dimensional high-order PCE}
		\label{fig:example1-mucoh}
	\end{subfigure}
	\begin{subfigure}[t]{0.45 \linewidth}
		\includegraphics[width=1\linewidth]{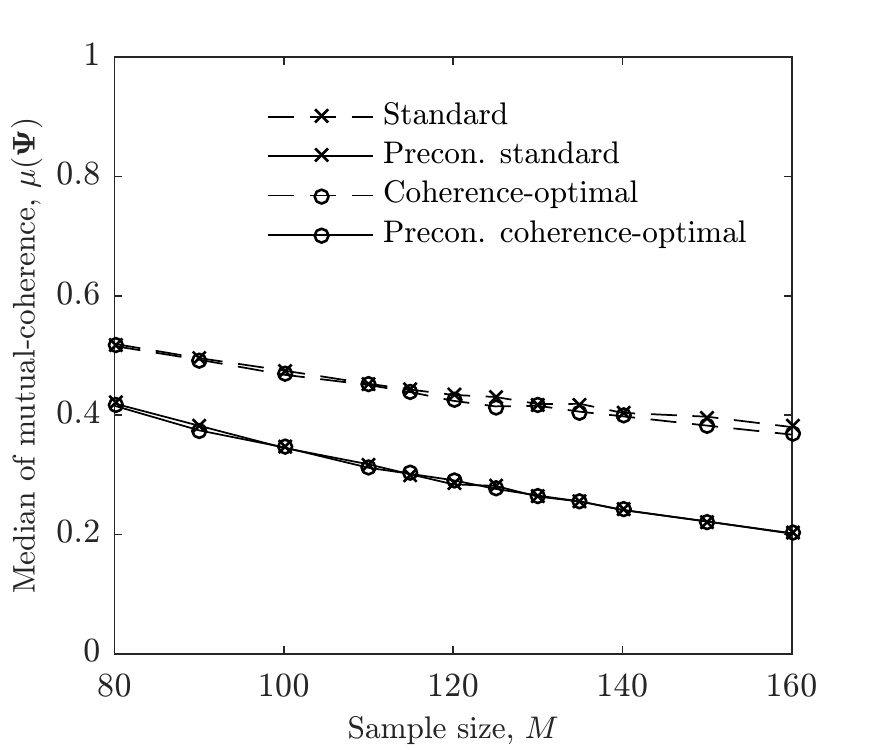}
		\caption{High-dimensional low-order PCE}
		\label{fig:example2-mucoh}
	\end{subfigure} \centering
	\caption{Comparison of mutual-coherence of measurement matrix  with and without preconditioning for standard and coherence-optimal sampling for (a) low-dimensional high-order manufactured PCE of Eq.~\ref{ex.lowDhighk} (b) high-dimensional low-order manufactured PCE of Eq.~\ref{ex.highDlowk} }
	\label{fig.example1and2-mucoh}
\end{figure}

Accuracy improvements observed in Figures \ref{fig.example1} and \ref{fig.example2} are direct results of improving the incoherence properties of measurement matrix. To demonstrate these improvements, Figures \ref{fig:example1-mucoh} and \ref{fig:example2-mucoh} compare the median of mutual-coherence of (equivalent) measurement matrix for all four approaches for low-dimensional high-order problem and high-dimensional low-order problem, respectively. It can be seen that for low-dimensional high-order problem, coherence-optimal sampling improves mutual-coherence more significantly than preconditioning alone. However, combination of coherence-optimal sampling and preconditioning leads to the smallest mutual-coherence. For high-dimensional low-order problem, it can be seen that coherence-optimal sampling slightly improves mutual-coherence, while on the other hand, the preconditioning scheme significantly improves the mutual-coherence. 

\subsection{An approximately sparse function}
\begin{figure}[H]
\centering
		\includegraphics[width=.4\linewidth]{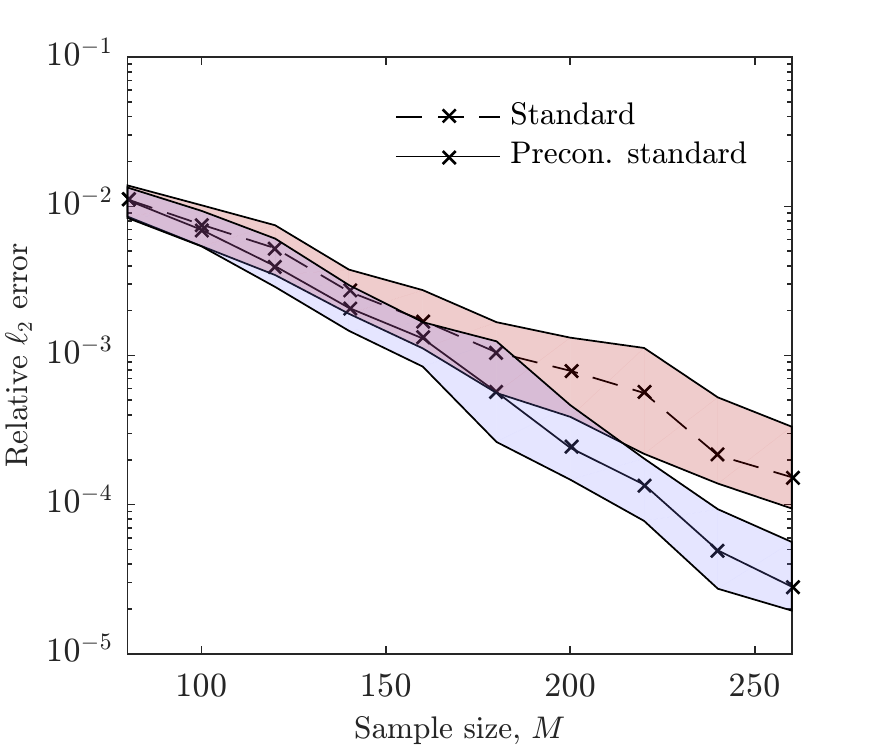}
		\caption{ Comparison of relative $\ell_2$ errors for sparse recovery of response with and without preconditioning for  the stochastic mass-spring problem } 
				\label{fig.Spring-error}
\end{figure}

In this section, we consider the  response of a mass-spring system as an approximately sparse target function. In particular, we   consider the stochastic mass-spring problem, given by
\begin{equation}\label{mass-spring}
m\frac{\mathrm{d}^2 x}{\mathrm{dt}^2}(t,\bm \Xi)+\gamma  x= f \text{sin}(\omega t),
\end{equation}
subject to initial conditions
\begin{equation*}
x(0)=0, \quad \dot{x}(0)=0.
\end{equation*}
We consider the mass $m$, spring constant $\gamma $ and frequency $\omega$ to be uncertain. We define $\bm \Xi=(m,\gamma ,\omega)$, where $m\in[0.018,0.022]$, $\gamma  \in[0.045,0.035]$ and $\omega \in[1.01,0.99]$. We choose $x(t=20)$ to be our QoI and use a 10th-order Legendre polynomial expansion to approximate the QoI. We employ standard sampling approach with and without preconditioning to estimate the coefficients of expansion, and use the analytical solution of Equation \ref{mass-spring} as the target function.  To compare the performance of standard sampling with and without preconditioning on this target function, the numerical results were obtained under a setting similar to that in the previous example with 100 independent runs. The cardinality of basis set is 286 and we aim to recover the vector of coefficients using a small set of samples. Figure \ref{fig.Spring-error} compares the relative error for the QoI of the mass-spring problem with and without preconditioning. It can be seen that approximation accuracy can be significantly improved using preconditioning.     

\subsection{An exactly sparse function with noisy measurement}
\begin{figure}[H]
	\begin{subfigure}[t]{0.33 \linewidth}
		\includegraphics[width=1\linewidth]{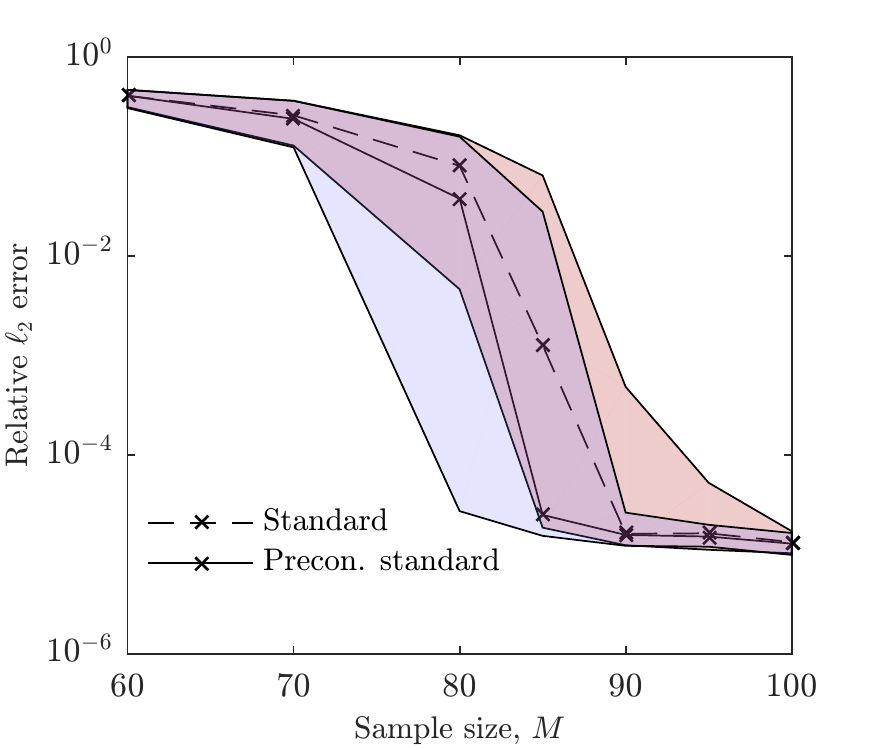}
		\caption{$\text{Measurement  noise}\sim N(0,10^{-3})$}
		\label{fig:Rosenbrock-error-3}		
	\end{subfigure}
	\begin{subfigure}[t]{0.33 \linewidth}
		\includegraphics[width=1\linewidth]{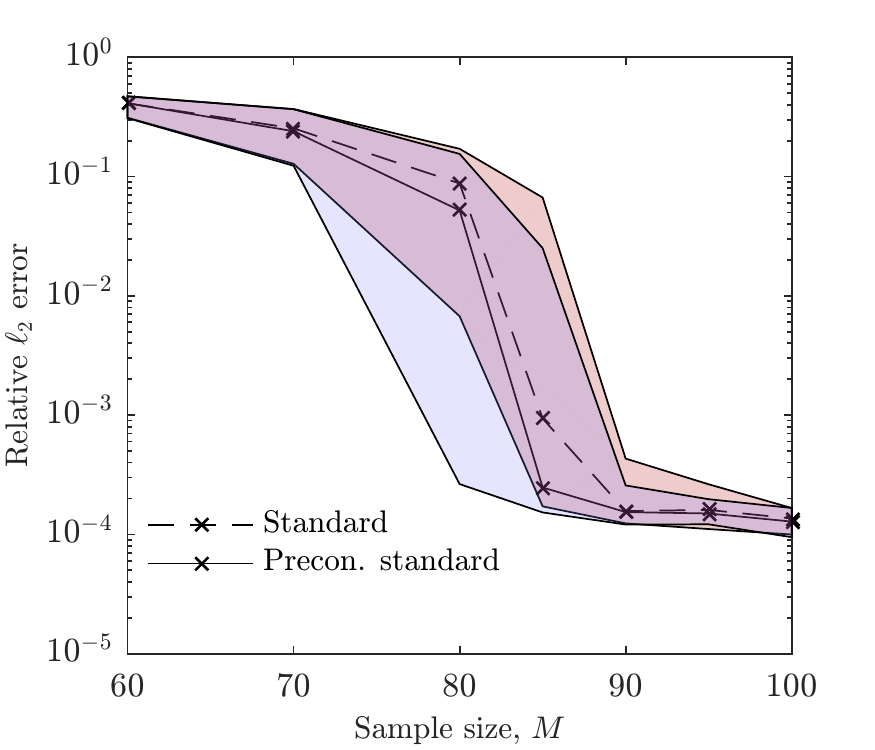}
		\caption{$\text{Measurement noise}\sim N(0,10^{-2})$ }
		\label{fig:Rosenbrock-error-2}
	\end{subfigure}
	\begin{subfigure}[t]{0.33 \linewidth}
		\includegraphics[width=1\linewidth]{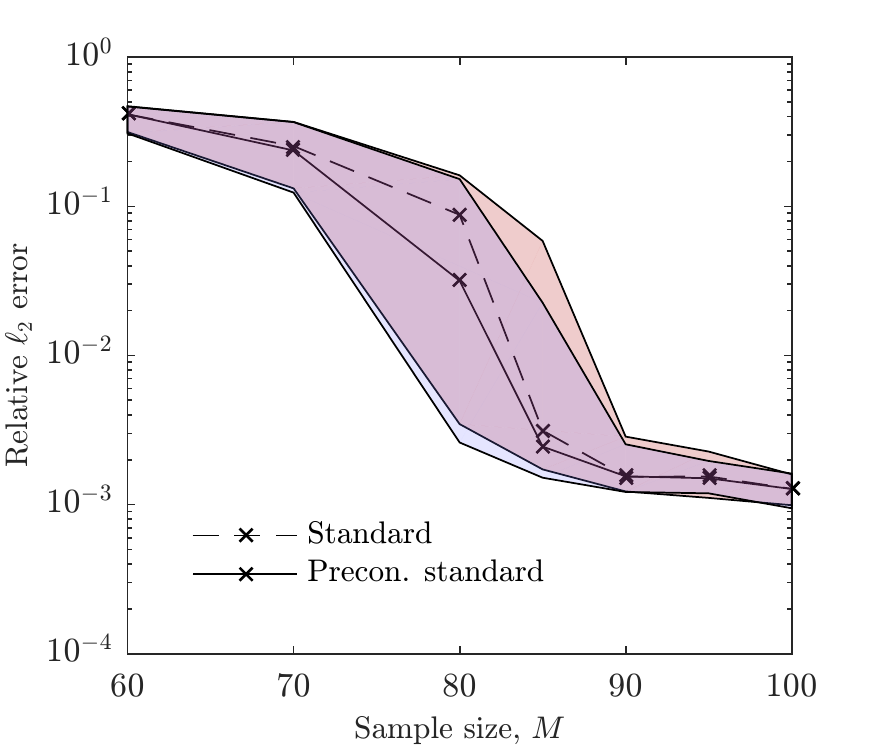}
		\caption{$\text{Measurement noise}\sim N(0,10^{-1})$}
		\label{fig:Rosenbrock-error-1}
	\end{subfigure}	
	\caption{ Comparison of relative errors in recovering the Rosenbrock function with noisy measurements by compressive sampling (using standard samples) with and without preconditioning.  }
	\label{fig.Rosenbrock}
\end{figure}

In  all the previous examples, measurements were noiseless.  Here, we  consider the measurements to be noisy in a problem with a ``moderate" combination of dimension and order. Specifically, consider the following 6-dimensional 4th-order generalized Rosenbrock function with random inputs following a uniform density on $\left [ -1,1 \right ]^6$,
\begin{equation}\label{ex.rosenbrock}
u(\bm \Xi)= \sum_{i=1}^{5} 100(\Xi_{i+1}-\Xi_i^2)^2+(1-\Xi_i)^2.
\end{equation}
We consider the measurement noise to follow a normal distribution with zero mean. For a better understanding of accuracy improvement, we consider three different standard deviations for the measurement noise: $10^{-3}, 10^{-2}$, and $10^{-1}$. Our objective is to recover the sparse vector of coefficients  using a sample set whose size is smaller than the cardinality of basis set which is 210. We report the results for 100 independent trials for standard sampling with and without preconditioning. It can be seen in  Figure \ref{fig.Rosenbrock}  that preconditioning significantly improves the approximation accuracy for standard sampling when the standard deviation of noise is relatively small. On the other hand, when standard deviation of noise is relatively large, it is more likely that the preconditioning matrix undesirably amplifies the noise vector. In this case, we observed that the cross-validation algorithm typically identified a larger $\lambda$ value, which limited the accuracy improvement that can be achieved by preconditioning. 

\section{Conclusion}

In this work, we introduced a preconditioning approach to improve the accuracy of compressive sampling-based recovery  of  polynomial chaos expansions.  We demonstrated the potential accuracy improvement offered solely by preconditioning a measurement matrix that has already been formed, e.g. in cases where samples are already collected. To do this, we multiply both sides of the equation system by a preconditioning matrix. The preconditioning matrix is designed such that the preconditioned measurement matrix has better incoherence properties and signal-to-noise ratio for the preconditioned problem is not undesirably large, thereby enhancing the recovery accuracy. We provided theoretical motivation for such scheme along with various numerical examples to validate the preconditioning approach.

\section{References}
\bibliography{bibfile}
%\bibliography{full_ref}

\end{document}